\documentstyle[12pt,epsfig]{article}
\newcommand\beq{\begin{equation}}
\newcommand\eeq{\end{equation}}
\newcommand\bea{\begin{eqnarray}}
\newcommand\eea{\end{eqnarray}}

\begin{document}
\vspace{-2.0cm}
\bigskip

\baselineskip=30pt
\centerline{\Large \bf SU(N) Coherent States} 
\vskip .8 true cm

\begin{center} 
{\bf Manu Mathur}\footnote{manu@bose.res.in} and
{\bf H. S. Mani} \footnote{hsmani@bose.res.in}
\vskip 0.5 true cm

S. N. Bose National Centre for Basic Sciences \\ 
JD Block, Sector III, Salt Lake City, Calcutta 700098, India

\end{center} 
\bigskip

\centerline{\bf Abstract}

\noindent We generalize Schwinger boson representation of SU(2) algebra to 
SU(N) and define coherent states of $SU(N)$ using $2(2^{N-1}-1)$ bosonic 
harmonic oscillator creation and annihilation operators. We  give an 
explicit construction of all (N-1) 
Casimirs of SU(N) in terms of these creation and annihilation operators.   
The SU(N) coherent states belonging to any irreducible representations 
of SU(N) are labelled by the eigenvalues of the Casimir operators
and are characterized  by (N-1) complex orthonormal 
vectors describing the SU(N) manifold.
The coherent states provide a resolution of identity, satisfy the continuity 
property, and possess a variety of group theoretic properties. 

\vskip .9 true cm
\noindent PACS: ~02.20.-a 
\vskip .4 true cm

\section{\bf Introduction}
The idea of coherent state for a quantum system was realized 
by Schr{\"o}dinger \cite{sch} way back in 1926 in the context of 
quantum state of classical motion for a harmonic oscillators.
This simplest coherent state construction is associated with 
the Heisenberg-Weyl group whose Lie algebra is given in terms 
of a harmonic oscillator creation and annihilation operators. 
These states have been widely used in physics \cite{glauber,klauder,
sudarshan}. 
Later, coherent states associated with arbitrary Lie group 
in a given representation were constructed \cite{perelomov} by 
applying group operators to a weight vector in that 
particular representation. In the special case of SU(2)
group, another equivalent way of constructing the coherent states 
was by exploiting the Schwinger representation of the SU(2) algebra
\cite{schwinger}. This representation involves 
a doublet of harmonic oscillator creation annihilation operators
in terms of which one generalizes the simplest Heisenberg-
Weyl coherent state construction to the SU(2) group. 
In \cite{md}, using the Schwinger representation of SU(3) 
algebra, we had constructed coherent states belonging to an 
arbitrary representation of SU(3).
The motivation of the present work is: a) to further generalise this 
Schwinger representation of SU(3) to SU(N) Lie algebra for arbitrary 
N,~~ b) exploit it to construct coherent states belonging to arbitrary 
irreducible representations of SU(N).  We also give an explicit 
characterization of the SU(N) coherent states in terms of (N-1) complex 
N-plets describing the SU(N) manifold. 
In the work \cite{nemoto}, SU(N) coherent states were constructed 
by applying the standard procedure of applying SU(N) group operator 
on the highest weight state of a SU(N) representation. 

\noindent The organization of the 
paper is as follows. In section 2 we will briefly describe the Heisenberg-
Weyl and  SU(2) coherent states in terms of harmonic oscillators. 
The section 2 is not only for the sake of completeness 
but also for setting up notation and language in a simpler 
setting before dealing with the larger SU(N) groups. 
This section also  emphasize the common spirit between them and 
our SU(N) coherent state formalism.  However,  
these groups being too simple, many features 
of larger SU(N) groups become redundant and hence this section 
fails to bring out a technique which can be generalized 
to SU(N).  Therefore, in section 3, we briefly mention the SU(3) coherent 
state construction in a framework which is equivalent to the one in 
\cite{md}  but can be easily generalized to SU(N). In section 4, we 
generalize this SU(3) procedure and explicitly construct SU(N) coherent 
states for arbitrary N. 

\section{\bf Heisenberg-Weyl and $SU(2)$ Coherent States}

\indent
Given a group $G$ and its manifold ${\cal M}$, the coherent states in a 
given representation $R$ are functions of $q$ parameters denoted by 
$\lbrace z^1,z^2,...z^q \rbrace$ describing ${\cal M}$, and are defined 
as
\beq
\vert \vec{z} >_{R} ~\equiv ~T_R \big(g(\vec{z})\big) ~\vert 0 >_R ~.
\label{gp} 
\eeq
Here $T_{R}\big(g(\vec{z})\big)$ is a group element in 
the representation $R$, and $\vert 0>_{R}$ is a fixed vector belonging to 
$R$. In the simplest example of the Heisenberg-Weyl group, 
the Lie algebra contains three generators. It is defined in terms 
of creation annihilation operators $(a, a^{\dagger})$ satisfying
\beq
[a,a^{\dagger}]= {\cal I}, ~~~[a,{\cal I}] =0, ~~~[a^{\dagger}, {\cal I}]=0 ~.
\eeq
This algebra has only one infinite dimensional irreducible 
representation which 
can be characterized by occupation number states $\vert n> \equiv 
{(a^{\dagger})^n \over \sqrt {n!}} \vert 0>$ with $n=0, 1, 2 ...$.
A generic group element in (\ref{gp}) can be characterized by 
$T\big(g\big) = \exp ~(i\alpha {\cal I} + z a^{\dagger} - \bar{z}a)$ 
with an angle $\alpha$ and a complex parameter $z$. Therefore, 
\bea
\vert \alpha,z>_\infty ~&=&~ \exp (i\alpha) ~\vert z>_{\infty} , \nonumber \\
\vert z>_{\infty}  ~&=&~ \exp (z a^{\dagger} - \bar{z} a) ~\vert 0>
= \sum_{n=0}^{\infty} F_n (z) ~\vert n> ~,
\label{wcs} 
\eea
where the sum runs over all the basis vectors of the infinite dimensional 
representation, and 
\beq
F_n (z) ~=~ {z^n \over \sqrt{n!}} ~\exp (-|z|^2 /2)
\label{co1} 
\eeq
are the coherent state expansion coefficients. This feature, i.e., an 
expansion of the coherent states in terms of basis vectors of a given 
representation with analytic functions of complex variables ($F_n (z)$) as 
coefficients, will also be present in the case of $SU(N)$ groups. 
It is easy to see that Eq. (\ref{wcs}) provides a resolution of identity 
with the measure $d\mu(z) = dzd{\bar{z}}$. 

\noindent We now briefly review the next simplest example, i.e., the Schwinger 
representation of SU(2) Lie algebra and the associated coherent states.  
The Lie algebra is 
\bea 
[J^{a},J^{b}] = i \epsilon^{abc} J^{c}
\label{su2} 
\eea
The algebra (\ref{su2}) can be realized in terms of 
a doublet of harmonic oscillator creation and annihilation operators 
${a} \equiv (a^1, a^2)$ and $ \vec{a}^{\dagger} \equiv (a^{\dagger}_1, 
a^{\dagger}_2)$ respectively \cite{schwinger}. 
They satisfy the simpler bosonic commutation relation
$[a^i,a^{\dagger}_j] = \delta^{i}_{~j}$ with $i,j =1,2$. The vacuum state 
is $\vert 0,0 >$. In terms of these operators, 
\beq
Q^a ~\equiv ~\frac{1}{2} ~a^{\dagger}_i ~(\sigma^a )^{i}_{~j} ~a^j ~,
\label{sch} 
\eeq
where $\sigma^a$ denote the Pauli matrices. (We will generally use the 
convention that repeated indices are summed over). It is easy to check that 
the operators in (\ref{sch}) satisfy the $SU(2)$ Lie algebra 
\bea
[Q^{a},Q^{b}] = i \epsilon^{abc} Q^{c}
\label{mtv} 
\eea
\noindent We also note that 
\bea
[Q^{a},a^{\dagger}_{i}] = a^{\dagger}_j {1 \over 2}({\sigma^{a}})^{j}_{~i} 
\label{su2trans}
\eea
\noindent The equation (\ref{su2trans}) implies that $(a^{\dagger}_{1}, 
a^{\dagger}_{2})$ transform like a SU(2) doublet. This fundamental 
representation of SU(2) will be  denoted by the Young diagram 
$\Sigma_{[1]}$, which is a single box. Therefore, we can realize 
all SU(2) irreducible representation on the Hilbert space of 
harmonic oscillators created by creation operators acting on the vacuum 
which is a direct product of the vacuum states for $a_{1}$ and $a_{2}$. 
The SU(2) Casimir operator is given by: 
\bea  
{\cal C} = \sum_{i=1}^{3} {a}^{\dagger}_{i} \cdot {a}^{i} 
\equiv {a}^{\dagger}\cdot{a} 
\label{su2cas} 
\eea 
\noindent The eigenvalues of the Casimir ${\cal C}$ will be 
denoted by C. The various irreducible representations of 
SU(2) are characterized by the eigenvalues of the Casimir (or occupation 
number operator) in (\ref{su2cas}). We also know that any irreducible 
representation of SU(2) can also be defined by its Young diagram which 
is obtained by arranging certain number of boxes in a row. It is 
easy to see that ${\cal C}$ just counts this number. The basis vectors 
of the SU(2) irreducible representation with ${\cal C} = C$ are 
given by: 
\bea
\Vert i_{1}i_{2}....i_{C}> \equiv a^{\dagger}_{i_{1}}a^{\dagger}_{i_{2}}
.....a^{\dagger}_{i_{C}} |0>  
\label{su2irrep} 
\eea 
\noindent The dimension of the irreducible representation above 
is (C+1). With the harmonic oscillator
creation and annihilation operators, $SU(2)$ coherent 
states can be  obtained by directly generalizing (\ref{wcs}). We 
define a doublet 
of complex numbers $(z^1, z^2)$ with the constraint:
\bea
|z|^{2} \equiv |z^1|^2 + |z^2|^2 =1 
\label{su2cons} 
\eea
The above constraint gives $3$ independent real compact parameters which 
define the sphere ${\cal M}=S^3$. 
The $SU(2)$ coherent state in the representation $C$ is now defined as
\bea
\vert z>_{C} \equiv \vert z^1, z^2>_{C} & ~=~ &  
\exp \Big( \vec{z} \cdot \vec{a}^\dagger 
\Big) ~\vert 0,0> 
~|_{{\vec{a}^{\dagger} \cdot \vec{a} = C}}.
\nonumber \\
& =~ & F^{i_{1}i_{2}...i_{C}} a^{\dagger}_{i_{1}}a^{\dagger}_{i_{2}}..
a^{\dagger}_{i_{C}}|0> \nonumber \\
&=& F^{i_{1}i_{2}...i_{C}}\Vert i_{1}i_{2}....i_{C}>  
\label{cs2} 
\eea
\noindent where, 
\bea
 F^{i_{1}i_{2}...i_{C}} = \frac{1}{C!} z^{i_{1}} z^{i_{2}}... z^{i_{C}} 
\label{chow} 
\eea
\noindent Note that F in (\ref{chow}) are analytic functions of $z^{1}$ 
and $z^{2}$.  Under SU(2) transformations: 
\bea 
z^{i} \rightarrow {z^{\prime}}^{i} = expi (\theta^{a} \frac{\sigma^{a}}{2})^{i}_
{~j}z^{j}
\label{uut}
\eea
\noindent Thus the constraint (\ref{su2cons}) remains invariant under the 
SU(2) transformations (\ref{uut}). Therefore, the coherent states $|z> 
\equiv |z_{1},z_{2}>$ transform amongst themselves on $S^{3}$.
It is easy to check the resolution of identity: 
\beq
\int ~d^{2}z^{1}d^{2}z^{2} \delta(|z^{1}|^{2}+|z^{2}|^{2} - 1)|z >_C
{}_{C} < z | ~=~ \frac{1}{1+C} ~ \sum_{i_{1},..,i_{C}=1}^{2}
\Vert i_{1}i_{2}....i_{C}> < i_{1}i_{2}....i_{C} \Vert
\eeq

\section{\bf The Schwinger $SU(3)$ Representation} 

The basic technique  behind the construction of SU(N) coherent states 
is to generalize the Schwinger representation of SU(3) Lie algebra 
\cite{md} to SU(N) Lie algebra.  We will, therefore, briefly describe 
the construction of SU(3) Lie algebra in terms of harmonic oscillators 
here in a new framework which is equivalent to that in \cite{md} but is 
directly generalizable to SU(N).

\noindent The rank of the SU(3) group is 2 and therefore it has two 
fundamental representations: 
Any SU(3) irreducible representation can be built up from the 
two fundamental representations, a triplet $3$ and an anti-triplet 
$ \bar{3}$. The later is an antisymmetric combinations of the two triplets. 
The  two fundamental representations will be associated with 
the two (fundamental) Young diagrams $\Sigma_{[1,0]}$, for the 
triplet representation and $\Sigma_{[0,1]}$, for its conjugate 
(anti-triplet) representation. The Young diagram $\Sigma_{[1,0]}$ 
is one box and $\Sigma_{[1,0]}$ is two boxes arranged vertically. 
We will also label the fundamental representations by the Greek 
indices $[\alpha], [\beta]$ 
taking values 1 and 2.  The components of any irreducible tensor T 
transforming according 
to $\Sigma_{[1,0]} (\Sigma_{[0,1]})$ will be denoted by 
$T[\alpha=1]_{i_{1}} (T[\alpha=2]_{i_{1}i_{2}} \equiv
-T_{i_{1}i_{2}}, (i_{1},i_{2} =1,2,3))$.\footnote{The components of 
the complex conjugate of these tensors will be denoted by $T^{*}[1]^{j_{1}} 
T^{*}[2]^{j_{1}j_{2}}$.} 
Further, the 8 representation matrices corresponding to these two 
fundamental representations will be denoted by $\lambda^{a}[\alpha]$
\footnote{Note that $\lambda^{a}[2]^{i_{1}i_{2}}_{{\phantom{j_{1}j_{2}}} 
j_{1}j_{2}} \equiv  {1 \over 2} \epsilon^{i_{1}i_{2}k}\epsilon_{j_{1}j_{2}l} 
({-\lambda^{*a}}[1])^{~l}_{k}$.}. 
If $Q^{a}$ are the generators of SU(3) then under the SU(3) they transform as 
\bea
\big[Q^{a}, T[1]_{i_{1}}\big] &=& T[1]_{j_{1}} {\lambda^{a}[1]}^{j_{1}}
_{~~i_{1}}
\nonumber \\  
\big[Q^{a}, T[2]_{i_{1}i_{2}}\big] & =& T[2]_{j_{1}j_{2}}   
{\lambda^{a}}[2]^{j_{1}j_{2}}_{~~~~ i_{1}i_{2}} 
\label{trans} 
\eea
The matrices on the right hand side of (\ref{trans}) are the matrices 
belonging to the fundamental representations of SU(3) and satisfy 
the SU(3) Lie algebra:
\bea
\big[\lambda^{a}[1],\lambda^{b}[1]\big] = i f^{abc}\lambda^{c}[1] \nonumber \\ 
\big[\lambda^{a}[2],\lambda^{b}[2]\big] = i f^{abc}\lambda^{c}[2] 
\label{gell} 
\eea 
The direct product of these two tensors (representations) span the 
whole SU(3) representation space. Infact, any irreducible tensor can be 
obtained by taking direct products of $C_{1}$ of T[1] and $C_{2}$ of 
T[2] tensors. The corresponding Young diagram is represented by 
arranging $C_{2}$ of $\sigma[2]$  Young diagrams and $C_{1}$ of 
$\sigma[1]$ diagrams from left 
to right side by side.  Following Schwinger representation of 
of SU(2) Lie Algebra, we introduce two sets of creation-annihilation 
operators $a[\alpha], a^{\dagger}[\alpha], \alpha =1,2$. The set 
$a[\alpha=1]$ represents three annihilation operators denoted by 
$a^{i_{1}}[1] \equiv (a[1]^{1}, a[1]^{2}, a[1]^{3})$ and $a[\alpha=2]$ 
represents another set of three annihilation operators, 
$a[2]^{i_{1}i_{2}} = - a[2]^{i_{2}i_{1}}$ having three independent 
components $(a[2]^{12}, a[2]^{23}, a[2]^{31})$\footnote{In [4],
we had chosen $b_{1} = a[2]^{23}, b_{2} = a[2]^{31}, b_{3} = a[2]^{12}$}. 
We impose the following commutation relations on them:  
\bea
\hspace{-1.0cm} \big[a[1]^{i_{1}},a^{\dagger}[1]_{j_{1}}\big]  =  
\delta^{i_{1}}_{j_{1}} & , &  
\big[a^{i_{1}}[1],a^{i_{2}}[1]\big] = 0 
\nonumber \\ 
\big[a[2]^{i_{1}i_{2}},a^{\dagger}[2]_{j{1}j_{2}}\big]  = 
\delta^{i_{1}}_{j_{1}}
\delta^{i_{2}}_{j_{2}} - \delta^{i_{1}}_{j_{2}}\delta^{i_{2}}_{j_{1}} 
& , &  
\big[a[2]^{i_{1}i_{2}},a[2]^{j{1}j_{2}}\big] = 0 
\nonumber \\
\big[a[1]^{i_{1}},a[2]^{i_{2}i_{3}}\big] = 0 &,& 
\big[a^{\dagger}[1]_{i_{1}},a[2]^{i_{2}i_{3}}\big] = 0; 
\label{alg} 
\eea
\noindent We now define the SU(3) generators in the Hilbert space 
of Harmonic oscillators as: 
\bea
Q^{a} = a^{\dagger}[1]_{i_{1}}\lambda^{a}[1]^{i_{1}}_{~~i_{2}}a[1]^{i_{2}}
+ {1 \over 2!} a^{\dagger}[2]_{i_{1}i_{2}}\lambda^{a}[2]^{i_{1}i_{2}}_{
~~~~j_{1}j_{2}}
a[2]^{j_{1}j_{2}}
\label{SU3}
\eea
\noindent Using (\ref{alg}), it is easy to check that 
\bea 
[Q^{a},Q^{b}] = i f^{abc}Q^{c} 
\label{algsu3} 
\eea
\noindent Further, 
\bea
\big[Q^{a}, a^{\dagger}[1]_{i_{1}}\big] &=& a^{\dagger}[1]_{j_{1}} {\lambda^{a}[1]}^{j_{1}}
_{~~i_{1}}
\nonumber \\
\big[Q^{a}, a^{\dagger}[2]_{i_{1}i_{2}}\big] & =& a^{\dagger}[2]_{j_{1}j_{2}}
{\lambda^{a}}[2]^{j_{1}j_{2}}_{~~~~ i_{1}i_{2}}
\label{33bar}
\eea
The equation (\ref{33bar}) implies that $a^{\dagger}[1]$ and $a^{\dagger}[2]$
transform like $\Sigma_{[1,0]}$ and $\Sigma_{[0,1]}$ fundamental 
representations  respectively. 
Therefore,  all the irreducible 
representations of (\ref{algsu3}) can be realized on  the Hilbert 
space created by creation operators $a^{\dagger}[1]_{i_{1}}$ and 
${a^{\dagger}[2]}^{i_{1}i_{2}}$ acting on the vacuum state $|0>$ which 
is direct product of vacuum states associated with each of the six 
harmonic oscillators. 
Further, the two Casimir operators of SU(3) in this generalization of 
Schwinger representation are given by: 
\bea 
{\cal C}[1] &  = & \sum_{i_{1}=1}^{3} a^{\dagger}_{i_{1}}[1]a^{i_{1}}[1]  
\nonumber \\ 
{\cal C}[2] & = & \sum_{i_{1},i_{2} =1}^{3} 
a^{\dagger}_{i_{1}i_{2}}[2]a^{i_{1}i_{2}}[2]  
\label{casimir} 
\eea
The eigenvalues of the two Casimirs in (\ref{casimir})  will be 
denoted by $C_{1}$ and $C_{2}$ respectively. 

\subsection{\bf Irreducible Representations of SU(3)} 

The eigenvalues of Casimirs $(C_{1}, C_{2})$  characterize 
all the irreducible representations of SU(3). 
On the other hand, we can also define irreducible representations 
by its Young diagram which is built up by arranging certain numbers 
of $\Sigma_{[1,0]}$ followed by $\Sigma_{[0,1]}$  horizonally from 
right to left. It is easy to see that ${\cal{C}}_{1}$ and 
${\cal{C}}_{2}$ just count these numbers. 
Therefore, we will denote a general SU(3) Young diagram 
by $\Sigma_{[C_{1},C_{2}]}$ which contains $C_{1}$ of $\Sigma_{[1,0]}$ 
and $C_{2}$  of $\Sigma_{[2]}$ put together from from right to left. 
Therefore, a general basis vector in $\Sigma_{[C_{1},C_{2}]}$ 
can be written as: 
\bea 
\Vert i_{1}^{1}i_{1}^{2}....i_{1}^{C_{2}+C_{1}};i_{2}^{1}i_{2}^{2}..
i_{2}^{C_{2}}>_{\Sigma_{[C_{1},C_{2}]}}  
& \equiv &  \Big( e_{\Sigma_{[C_{1},C_{2}]}}\Big)  
a^{\dagger}[2]_{i^{1}_{1}i_{2}^{1}} a^{\dagger}[2]_{i^{2}_{1}i^{2}_{2}}.....
a^{\dagger}[2]_{i_{1}^{C_{2}}i_{2}^{C_{2}}}[2] \nonumber \\  
&& a^{\dagger}[1]_{i_{1}^{C_{2}+1}}a[1]^{\dagger}_{i_{1}^{C_{2}+2}}....
a^{\dagger}[1]_{i_{1}^{C_{2}+C_{1}}}|0> 
\label{irrep} 
\eea
In (\ref{irrep}), we have characterised the basis vectors of 
$\Sigma_{[C_{1},C_{2}]}$ representation by the various box indices 
appearing in the Young Tableau $\Sigma_{[C_{1},C_{2}]}$. More explicitly, 
the index $i_{r}^{c}$ represents the index corresponding to the box 
appearing in the $r^{th}$ row and $c^{th}$ column.     
$e_{\Sigma_{[C_{1},C_{2}]}}$ is the idempotent associated with 
he Young tableau $\Sigma_{[C_{1},C_{2}]}$. In general, 
it is an element of group algebra corresponding to the permutation group 
$S(C_{1}+2C_{2})$.  It is defined as the product of the two  
symmetrisers ${\cal{S}}_{1}$ (symmetrizing the indices in the 
first row $(i_{1}^{1},i_{1}^{2},..., i_{1}^{C_{1}+C_{2}})$, 
${\cal{S}}_{2}$ (symmetrizing the indices in the
second row $(i_{2}^{1},i_{2}^{2},..., i_{2}^{C_{1}+C_{2}})$
and $C_{2}$ antisymmetrisers (acting on the $C_{2}$ columns indices 
$\big((i^{1}_{1},i^{1}_{2}),(i^{2}_{1},i^{2}_{2}),..,
(i^{C_{2}}_{1},i^{C_{2}}_{2})\big)$. In fact, the idempotent for a 
particular Young tableau can be constructed in two different ways,
first symmetrizing the indices in each row and then antisymmetrizing 
the indices in each column or first antisymmetrizing the indices in 
each column followed by symmetrizing the indices of each row. Both 
procedure lead to equivalent results. Here, since the column 
indices are already anti-symmetric, we choose the latter alternative
and need to consider only the two symmetry operations along the two 
rows of $\Sigma_{[C_{1},C_{2}]}$ respectively\footnote{In (\ref{s1})
and (\ref{s2}),
the $p \in S_{n}$ is an element of the permutation group $S_{n}$ and 
it denotes the permutation $ \left( \begin{array}{ccc} 1 & 2 &.... N  \\
p_1 & p_2 &.... p_N \end{array}\right)$
element of the permutation group $S_{N}$. Henceforth, we follow this 
notation throughout the paper.}:   
\bea
{\cal{S}}[1]  = \sum_{p \in S_{C_{2}+C_{1}}} p 
\label{s1} 
\eea 
\bea
{\cal{S}}[2]  = \sum_{p \in S_{C_{2}}} p 
\label{s2} 
\eea 
The idempotent is just the product of  (\ref{s1}) and (\ref{s2})
and is given by: 
\bea
e_{\Sigma_{[C_{1},C_{2}]}} = {\cal{S}}[1]_{[i]} {\cal{S}}[2]_{[j]} 
\eea 
One can easily see that: 
\bea 
{\cal{C}}[1] |i_{1}^{1}i_{1}^{2}....i_{1}^{C_{2}+C_{1}};i_{2}^{1}i_{2}^{2}..
i_{2}^{C_{2}}> &=& C_{1} |i_{1}^{1}i_{1}^{2}....i_{1}^{C_{2}+C_{1}};
i_{2}^{1}i_{2}^{2}..i_{2}^{C_{2}}>  \nonumber \\
{\cal{C}}[2] |i_{1}^{1}i_{1}^{2}....i_{1}^{C_{2}+C_{1}};i_{2}^{1}i_{2}^{2}..
i_{2}^{C_{2}}> &=& C_{2} |i_{1}^{1}i_{1}^{2}....i_{1}^{C_{2}+C_{1}};
i_{2}^{1}i_{2}^{2}..i_{2}^{C_{2}}> 
\label{ic} 
\eea 
Thus all the irreducible representations are eigenvectors of 
the two casimirs ${\cal{C}}[1]$ and ${\cal{C}}[2]$. 
As an example we consider $C_{1} = C_{2} = 1$ corresponding 
to the octet representation of SU(3). For this representation, 
${\cal{S}}[1] = \big(e + (i^{1}_{1},i^{2}_{1})\big)$  and 
${\cal{S}}[2] = e$. Therefore,  (1,1) 
representation is given by: 
$\Vert i_{1}^{1},i_{1}^{2};i^{1}_{2}>_{\Sigma_{[1,1]}}  = 
\big[a^{\dagger}[2]_{i^{1}_{1}i^{1}_{2}}
a^{\dagger}[1]_{i^{2}_{1}} + a^{\dagger}[2]_{i^{2}_{1}i_{2}^{1}}
a^{\dagger}[1]_{i^{1}_{1}}\big]|0>$. 

\noindent In \cite{sharat} similar results as in this section have been 
obtained by using Bargmann's techniques for SU(3). More recently, SU(3) 
multiplicity problem has been analysed by exploiting the above Schwinger 
construction and a mutually commuting Sp(2,R) group \cite{mukunda}. 

\subsection{\bf The $SU(3)$ Coherent States}

We consider two triplets of complex numbers $z{[\alpha]} (= z^{1}{[\alpha]}, 
z^{2}{[\alpha]}, z^{3}{[\alpha]}), \alpha =1,2$ describing 
the Euclidean manifold $R^{6} \otimes R^{6}$. We impose the orthonormality 
conditions: 
\bea 
{\bar z}{[\alpha]}.z{[\beta]} \equiv \sum_{i=1}^{3} z^{*}_{i}[\alpha]
z^{i}[\beta] = \delta_{\alpha,\beta} 
\label{onc1}
\eea
\noindent We now define another vector $z[1,2]$ transforming as 
$\Sigma_{[0,1]}$ with components: 
\bea 
z^{i_{1}i_{2}}[12] \equiv \sqrt{1 \over 2} \left(z[1]^{i_{1}} z[2]^{i_{2}} 
- z[2]^{i_{1}}z[1]^{i_{1}}\right) 
 = \sqrt{1 \over 2}\epsilon^{\alpha_{1}\alpha_{2}}  z[\alpha_{1}]^{i_{1}} 
z[\alpha_{2}]^{i_{2}}  
\eea 
Using (\ref{onc1}), we find: 
\bea
\label{onc2}
z[1,2].z[1,2]^{*} & \equiv & z^{i_{1}i_{2}}[[12]{z^{*}}_{i_{1}i_{2}}[12] = 
1 \\  
\tilde{z}[12].z[1] \equiv  \epsilon_{i_{1}i_{2}i_{3}}z^{i_{1}i_{2}}[12]
z^{i_{3}}[1] = 0 &,&  
\tilde{z}[12].z[2] \equiv  \epsilon_{i_{1}i_{2}i_{3}}z^{i_{1}i_{2}}[12]
z^{i_{3}}[2] = 0 \nonumber
\eea
We can now construct any $3 \times 3$ SU(3) matrix ${\cal U}_{3}$ in 
terms of z[1] and  z[2]: 
\bea 
{\cal U}_{3} ~=~ \left( \begin{array}{ccc} z[1]_1 & z[1]_2 & z[1]_3  \\
z[2]_1 & z[2]_2 & z[2]_3 \\
\bar{{\tilde z}}[12]_1 & \bar{{\tilde z}}[12]_2 & \bar{\tilde z}[12]_3 
\end{array}\right)
\label{su3mat}
\eea 
Thus we immediately see that z[1] and z[2] with (\ref{onc1}) and 
(\ref{onc2}) completely describe the SU(3) manifold. 
At this stage we  define the SU(3) coherent states generating function as: 
\bea 
|z[1],z[2]>_{C_{1},C_{2}} \equiv exp\left(z[1]^{i_{1}}a^{\dagger}[1]_{i_{1}}
 + z[1,2]^{i_{1}i_{2}} a^{\dagger}[2]_{i_{1}i_{2}}\right) |0> 
{\mid}_{{}^{a^{\dagger}[1].a[1] = C_{1}}_{a^{\dagger}[2].a[2] = C_{2}}}
\label{cs33}
\eea 
\noindent One can see that in the expansion of (\ref{cs33}), due 
to the constraints (\ref{onc1}) and (\ref{onc2}), each 
irreducible representation occurs {\it once and only once}. 
We will now show that (\ref{cs33}) indeed generates all the 
coherent states of SU(3). 
The constraints on the right hand side of (\ref{cs33}), 
$(a^{\dagger}[1].a[1] = C_{1}$ and $a^{\dagger}[2].a[2] = C_{2})$, 
select different  possible irreducible representations of SU(3). 
Note that the states in (\ref{cs33}) are characterised by the continuous 
parameters $(z[1],z[2])$ on the SU(3) manifold.  It is easy to check the 
SU(3) transformation properties of z[1] and z[1,2], 
\bea 
z[1]^{i_{1}} & \rightarrow &  z^{\prime}[1]^{i_{1}} = \big(expi~\theta^{a} 
\lambda^{a}[1] \big)^{i_{1}}_{~i_{2}} z[1]^{i_{2}} \nonumber \\
z[1,2]^{i_{1}i_{2}} & \rightarrow &  z^{\prime}[12]^{i_{1}i_{2}}  = 
\big(expi ~\theta^{a} \lambda^{a}[12]
\big)^{i_{1}i_{2}}_{~~~~j_{1}j_{2}} z[1,2]^{j_{1}j_{2}} 
\label{transf} 
\eea 
\noindent 
In (\ref{transf}) $\theta^{a}$ are the 8 transformation associated 
with the SU(3) group transformation. Therefore, under SU(3) transformations 
both z[1] and z[2] transform like a triplet and  the orthonormality 
conditions (\ref{onc1}) and (\ref{onc2}) remain invariant under the SU(3) 
transformations and the state in (\ref{cs33}) defined at a point (z[1],z[2]) 
transform to the coherent state at $(z^{\prime}[1], z^{\prime}[2])$ on the 
SU(3) manifold. 

\noindent From the generating function (\ref{cs33}) we find: 
\bea 
|z[1],z[2]>_{C_{1},C_{2}}  = F^{i_{1}^{1}..i_{1}^{C_{1}+C_{2}}; 
i^{1}_{2}...i_{2}^{C_{2}}}
a[2]^{\dagger}_{i^{1}_{1}i^{1}_{2}}..a^{\dagger}[2]_{i_{1}^{C_{2}}
i_{2}^{C_{2}}} a^{\dagger}[1]_{i_{1}^{C_{2}+1}}.. 
a^{\dagger}[1]_{i_{1}^{C_{2}+C_{1}}}|0>  
\label{csn22} 
\eea
\noindent where, 
\bea 
C_{1}! C_{2}!  F(z[1],z[2])^{i_{1}^{1}..i_{1}^{C_{1}+C_{2}};
i^{1}_{2}...i_{2}^{C_{2}}}  & = & 
z[1,2]^{i^{1}_{1}i^{1}_{2}}z[1,2]^{i^{2}_{1}i^{2}_{2}}.......
z[1,2]^{i_{1}^{C_{2}} i_{2}^{C_{2}}}
\nonumber \\   
& & z[1]^{i_{1}^{C_{2}+1}}z[1]^{i_{1}^{C_{2}+2}}..z[1]^{i_{1}^{C_{2}+C_{1}}} \nonumber \\
&= & P_{2}~ z[1]^{i^{1}_{1}}z[1]^{i^{2}_{1}}...z[1]^{i_{1}^{C_{2}}} 
z[1]^{i_{1}^{C_{2}+1}}.. z[1]^{i_{1}^{C_{2}+C_{1}}} \nonumber \\
& & z[2]^{i_{2}^{1}}z[2]^{i_{2}^{2}}...z[2]^{i_{2}^{C_{2}}} 
\label{abc}
\eea 
\noindent In the second step in (\ref{abc}), we have used the 
anti-symmetry properties of $a^{\dagger}[2]^{'s}$ 
under the interchange of its two indices leading to $P_{2} = 
{(2!)}^{C_{2} \over 2}$.  The equation (\ref{abc}) 
clearly illustrates the following two important properties of the 
tensor $F\left(z[1],z[2]\right)$:  
\begin{enumerate}
\item It is analytic function of (z[1], z[2]) which describe the 
SU(3) manifold, 
\item It has exactly the same symmetry as that of $e_{\Sigma_{[C_{1},C_{2}]}}$, 
i.e 
\bea 
\hspace{-1.6cm} F\left(z[1],z[2]\right)^{i_{1}^{1}..i_{1}^{C_{1}+C_{2}};
i^{1}_{2}...i_{2}^{C_{2}}} 
= F\left(z[1],z[2]\right)^{i_{1}^{1}..i_{1}^{C_{1}+C_{2}};
i^{1}_{2}...i_{2}^{C_{2}}}
e_{\Sigma_{[C_{1},C_{2}]}} 
\eea 
\end{enumerate} 
Therefore, we can write (\ref{csn22}) as:  
\bea 
|z[1],z[2]>_{C_{1},C_{2}} & = &   
F^{i_{1}..i_{C_{1} + C_{2}}j_{1}...j_{C_{2}}} e_{\Sigma_{[C_{1},C_{2}]}} 
a[2]^{\dagger}_{i_{1}j_{1}}..a^{\dagger}[2]_{i_{C_{2}}j_{C_{2}}}  
a[1]^{\dagger}_{i_{C_{2}+1}}.. a[1]^{\dagger}_{i_{C_{2}+C_{1}}}|0>  
\nonumber \\ 
&= & 
F^{i_{1}..i_{C_{1} + C_{2}}j_{1}..j_{C_{2}}} \Vert i_{1}j_{1},i_{2}j_{2}..
i_{C_{2}}j_{C_{2}}
i_{C_{2}+1},i_{C_{2}+2}....i_{C_{2}+C_{1}}>_{[C_{1},C_{2}]}  
\eea
\noindent All these features are similar to Heisenberg-Weyl and SU(2) Schwinger 
coherent states of the previous section and are discussed in detail in 
\cite{md}.  

\noindent We now check the resolution of identity. The Haar measure on SU(3) 
manifold is given by: 
\bea
\int d\mu(z) \equiv \Big(\int\prod_{\alpha=1}^{2} dz[\alpha]\Big) \Big(
\prod_{\alpha,\beta} \delta(z[\alpha].z^{*}[\beta] -\delta_{\alpha,\beta})  
\eea 
We construct an operator ${\cal{O}}_{[3]}$: 
\bea
{\cal{O}}_{3} \equiv \int d\mu(z) |z[1],z[2]>_{{}_{C_{1},C_{2}}}
{}_{{}_{C_{1},C_{2}}}<z[1],z[2]|
\label{op} 
\eea
\noindent Under SU(3) transformations (\ref{transf}), ${\cal{O}}_{[3]}$ 
remains invariant. Therefore,
\bea
[Q^{a},{\cal{O}}_{[3]}] = 0, ~~~~ \forall a =1,2,....,8
\eea 
\noindent The Schur's Lemma implies: 
\bea 
{\cal{O}}_{[3]} = K I_{[C_{1},C_{2}]}
\label{idd}
\eea 
\noindent In (\ref{idd}), K is a constant and 
$I_{[C_{1},C_{2}]}$ is the identity operator in 
the $\Sigma_{[C_{1},C_{2}]}$  irreducible 
representation subspace.  Thus we finally get: 
\bea
\hspace{-1.5cm} & &\int d\mu(z[1],z[2]) |z[1],z[2]>_{{}_{C_{1},C_{2}}} 
{}_{{}_{C_{1},C_{2}}}<z[1],z[2]| = \int d^{6}z[1]d^{6}z[2] 
\prod_{\alpha,\beta=1}^{2} \delta\left(z^{*}[\alpha].z[\beta]-
\delta_{\alpha,\beta}\right) \nonumber \\  
\hspace{-1.5cm} &&F^{i_{1}..i_{C_{1} + C_{2}}j_{1}..j_{C_{2}}}
{F^{*}}_{k_{1}..k_{C_{1} + C_{2}}l_{1}..l_{C_{2}}} 
|{}_{i_{1}j_{1},.. i_{C_{2}}j_{C_{2}};
i_{C_{2}+1},..i_{C_{2}+C_{1}}}>_{{}_{[C_{1},C_{2}]}}
{}_{{}_{[C_{1},C_{2}]}}<{}^{k_{1}l_{1},..k_{C_{2}}l_{C_{2}};
k_{C_{2}+1},..k_{C_{2}+C_{1}}}|  \nonumber \\ 
&& = K |{}_{i_{1}j_{1},i_{2}j_{2},.. i_{C_{2}}j_{C_{2}};
i_{C_{2}+1},..i_{C_{2}+C_{1}}}>_{{}_{[C_{1},C_{2}]}}
{}_{{}_{[C_{1},C_{2}]}}<{}^{i_{1}j_{1},i_{2}j_{2},..i_{C_{2}}j_{C_{2}};
i_{C_{2}+1},..i_{C_{2}+C_{1}}}| \equiv K I_{[C_{1},C_{2}]}  
\label{id}
\eea 

\section{\bf The Schwinger $SU(N)$ Representation}

We now generalise the ideas developed in the previous section in 
the case of SU(3) to the group SU(N) for arbitrary N. The rank 
of the SU(N) group is (N-1). Therefore, there are (N-1) fundamental
representations denoted by the Young diagrams with 1,2 .... (N-1) 
vertical boxes respectively. Any irreducible representation of 
SU(N) can be built by taking direct products of the above (N-1) 
fundamental representations. Following SU(2) and SU(3), 
we introduce  (N-1) sets of creation annihilations operators 
$\big(a[1]^{i_{1}},a^{\dagger}[1]_{i_{1}}\big); \big(a[2]^{i_{1}i_{2}},
a^{\dagger}[2]_{i_{1}i_{2}}\big); ....; \big(a[N-1]^{i_{1}i_{2}.....i_{N-1}},
a^{\dagger}[N-1]_{i_{1}i_{2}.....i_{N-1}}\big)$, which can 
be written in a compact form as $(a[\alpha]^{i_{1}i_{2}.....i_{\alpha}}, \\
a^{\dagger}[\alpha]_{i_{1}i_{2}.....i_{\alpha}}), \alpha =1,2....(N-1)$.  
The commutation relations are straightforward generalization of 
(\ref{alg}) and are given by: 
\bea 
\big[a[\alpha]^{i_{1}i_{2}......i_{\alpha}},a^{\dagger}[\beta]_{j_{1}j_{2}......
j_{\beta}}\big]
& = & \delta_{\alpha,\beta} \sum_{p \in S_{\alpha}} 
(-)^{|p|} \delta^{~~i_{1}}_{j_{p_{1}}} 
\delta^{~~i_{2}}_{j_{p_{2}}}....\delta^{~~i_{\alpha}}_{j_{p_{\alpha}}} \nonumber \\
\big[a[\alpha]^{i_{1}i_{2}......i_{\alpha}},a[\beta]^{j_{1}j_{2}......j_{\beta}}\big] 
 =  0 &,&  
\big[a^{\dagger}[\alpha]_{i_{1}i_{2}......i_{\alpha}},
a^{\dagger}[\beta]_{j_{1}j_{2}...... j_{\beta}}\big]  =  0. 
\label{algn}
\eea 
\noindent In (\ref{algn}), $|p|=0$ if p is an even permutation and 
$|p| = 1$ if it is an odd permutation. 
We denote the (N-1) generators  belonging to (N-1) fundamental 
representation of SU(N) by $\lambda^{a}[\alpha]$, $a=1,2,..,(N^{2}-1)$. 
They satisfy the SU(N) 
algebra: 
\bea
\big[\lambda^{a}[\alpha],\lambda^{b}[\alpha]\big] = i f^{abc} 
\lambda^{c}[\alpha]~~; ~~~~\alpha =1,2....(N-1)
\label{mat}
\eea 
In(\ref{mat}) $f^{abc}$ are the SU(N) structure constants. 
The SU(3) generators in terms of Harmonic oscillators 
are given by: 
\bea
Q^{a} = a^{\dagger}[1]_{i_{1}}[1]\lambda^{a}[1]^{i_{1}}_{~~i_{2}}
a[1]^{i_{2}}    
+ {1 \over 2!}a^{\dagger}[2]_{i_{1}i_{2}}
\lambda^{a}[2]^{i_{1}i_{2}}_{~~~~j_{1}j_{2}}
a[2]^{j_{1}j_{2}} + .........\nonumber \\
..... + {1 \over (N-1)!} a^{\dagger}[N-1]_{i_{1}i_{2}..i_{N-1}}
\lambda^{a}[N-1]^{i_{1}..i_{N-1}}_{~~~~~~j_{1}..j_{N-1}} 
a[N-1]^{j_{1}j_{2}...j_{N-1}} \nonumber \\
= \sum_{\alpha=1}^{N-1} {1 \over \alpha !} a^{\dagger}[\alpha]_{i_{1},i_{2}...i_{\alpha}}
{\lambda^{a}}[\alpha]^{i_{1}i_{2}...i_{\alpha}}_{~~~~~~j_{1}j_{2}...j_{\alpha}}
a[\alpha]^{j_{1}j_{2}....j_{\alpha}} \hspace{3cm} 
\label{sun}
\eea
The commutation relations (\ref{algn}) imply: 
\bea
\big[Q^{a},Q^{b}\big] = i f^{abc} Q^{c}
\eea 

\noindent It is easy to verify that under SU(3) transformations the 
various creation operators belonging to the ${\alpha}^{th}$ representation 
transform amongst themselves as: 
\bea
\Big[Q^{a},a^{\dagger}_{i_{1}i_{2}....i_{\alpha}}[\alpha]\Big] = 
a^{\dagger}_{j_{1}j_{2}....j_{\alpha}}[\alpha] ({\lambda^{a}})^{j_{1}j_{2}...
j_{\alpha}}_{~~~~~~~i_{1}i_{2}...i_{\alpha}}[\alpha]
\label{fandu} 
\eea

\noindent This implies that all the irreducible representations of SU(N) 
can be realised on the Hilbert space created by the above creation 
operators acting on the vacuum.  
Further, the (N-1) Casimir operators are just the (N-1) number operators 
corresponding to (N-1) types of creation annihilation operators and are  
given by:
\bea
{\cal C}[1] &  = & \sum_{i_{1}=1}^{N} a^{\dagger}[1]_{i_{1}}a[1]^{i_{1}}
\nonumber \\
{\cal C}[2] & = & \frac{1}{2!} \sum_{i_{1},i_{2} =1}^{N}
a^{\dagger}[2]_{i_{1}i_{2}}a[2]^{i_{1}i_{2}} \nonumber \\
&.....&\nonumber \\
{\cal C}[N-1] & = & \frac{1}{(N-1)!}\sum_{i_{1},i_{2},....,i_{N-1} =1}^{N}
a^{\dagger}[N-1]_{i_{1}i_{2}....i_{N-1}}a[N-1]^{i_{1}i_{2}....i_{N-1}} 
\label{casimirn}
\eea
It is easy to check that the (N-1) Casimir operators in (\ref{casimirn}) 
commute with all the $(N^{2}-1)$ SU(N) generators in (\ref{sun}).  

\subsection{\bf The Irreducible Representations of SU(N)} 

All irreducible representations of SU(N) can be constructed 
by taking direct products of the (N-1) fundamental representations 
and then applying appropriate projection operators. Like the SU(3) 
case in the previous section, we now consider the most general SU(N) 
irreducible representation $\Sigma_{[C_{1},C_{2}......C_{N-1}]}$ containing 
$C_{\alpha}$ copies of the ${\alpha}^{th}$ fundamental representations. 
We formally writei\footnote{The explicit form of these basis vectors are 
given at the end of the next section (\ref{csf11}) and (\ref{irrepnn})}: 
\bea 
|C_{1},C_{2},.., C_{N-1}> \equiv 
e_{\Sigma_{[C_{1}...{C_{N-1}}]}}\big(a^{\dagger}[1]\big)^{C_{1}} 
\big(a^{\dagger}[2]\big)^{C_{2}} 
\big(a^{\dagger}[3]\big)^{C_{3}}.... 
\big(a^{\dagger}[N-1]\big)^{C_{N-1}}|0>.   
\label{irrepn} 
\eea
Just as in the case of SU(3) in the previous sections, all  
the creation operators being completely anti-symmetric in their 
indices,  the idempotent 
$e_{\Sigma_{[C_{1}..C_{N-1}}]}$ in (\ref{irrepn}) is 
constructed by multiplying all (N-1) symmetrizers: 
\bea 
e_{\Sigma_{[C_{1}...{C_{N-1}}]}} = {\cal{S}}_{1}{\cal{S}}_{2}...{\cal{S}}_{N-1}
\label{idemp} 
\eea 
\noindent where ${\cal{S}}_{1}, {\cal{S}}_{2}...{\cal{S}}_{N-1}$ are the 
elements of permutation group algebras associated with the permutation 
groups $S_{C_{1}+..+C_{N-1}}, S_{C_{2}+..+C_{N-1}},...,S_{C_{N-1}}$
respectively, 
\bea
{\cal{S}}_{1} &  \equiv & {1 \over (C_{1}+...+C_{N-1})!} 
\sum_{p \in S_{C_{1}+..+C_{N-1}}} p \nonumber \\
{\cal{S}}_{2} & \equiv & {1 \over (C_{2}+...+C_{N-1})!}
\sum_{p \in S_{C_{2}+..+C_{N-1}}} p \nonumber \\
&.....& \nonumber  \\
{\cal{S}}_{N-1} & \equiv & {1 \over (C_{N-1})!}
\sum_{p \in S_{C_{N-1}}} p 
\label{symm} 
\eea 
Again, as in the case of SU(3), it is easy to verify: 
\bea 
{\cal{C}}_{\alpha}|C_{1},C_{2},.., C_{N-1}>  = C_{\alpha} 
|C_{1},C_{2},.., C_{N-1}> 
\eea 

Therefore, the Casimir operator ${\cal{C_{\alpha}}}$ acting on an 
irreducible representation generated by the basis vectors 
${\cal{C}}_{\alpha}|C_{1},C_{2},.., C_{N-1}>$ just counts the number 
of times the ${\alpha}^{th}$ fundamental representation appears in it. 
This is again similar to the SU(3) case in the previous section. 
We now exploit this feature to construct the SU(N) coherent states. 
 
\subsection{The SU(N) Coherent States} 
We consider (N-1) N-plets of complex numbers $z{[\alpha]} (= z^{1}{[\alpha]}, 
z^{2}{[\alpha]},... z^{N}{[\alpha]}), \alpha =1,2,..,N-1$ describing 
the Eucledian manifold which is a direct product of (N-1) $R^{2N}$
We impose the orthonormality conditions: 
\bea 
{z^{*}}{[\alpha]}.z{[\beta]} = \delta_{\alpha,\beta} 
\label{onc}
\eea
We now define another vector $z[1,N-1]$ with components: 
\bea 
z^{i_{1}i_{2}..i_{N-1}}[1,N-1] & \equiv & \sqrt{1 \over (N-1)!} 
\sum_{p \in S_{N-1}} (-)^{|p|} 
\left(z[1]^{i_{p_{1}}} z[2]^{i_{p_{2}}}....z[N-1]^{i_{p_{N-1}}}\right) 
\nonumber \\
& = & \sqrt{1 \over (N-1)!}\epsilon^{\alpha_{1}\alpha_{2} ....{\alpha}_{N-1}}  
z[\alpha_{1}]^{i_{1}} z[\alpha_{2}]^{i_{2}} ....z[\alpha_{N-1}] 
\label{lly} 
\eea 
\noindent Using (\ref{onc}), we find: 
\bea
z[1,N-1].z[1,N-1]^{*} \equiv z^{i_{1}i_{2}..i_{N-1}}[[1,N-1]
{z^{*}}_{i_{1}i_{2}..i_{N-1}}[1,N-1] = 1  
\nonumber  \\
\tilde{z}[1,N-1].z[\alpha] \equiv  \epsilon_{i_{1}i_{2}....i_{N}}
z^{i_{1}i_{2}....i_{N-1}}[1,N-1] z^{i_{N}}[\alpha] = 0. 
\label{oncn}
\eea
We can now construct any $N \times N$ SU(N) matrix ${\cal U}_{N}$ in terms of 
z[1], z[2],......,z[N-1]: 
\bea 
{\cal U}_{N}  ~=~ \left( \begin{array}{ccc} z[1]^1 & z[1]^2 &.... z[1]^N  \\
z[2]^1 & z[2]^2 &.... z[2]^N  \\
: \\
:\\
z[N-1]^1 & z[N-1]^2 &.... z[N-1]^N  \\
{\bar{\tilde z}}[1..N-1]^1 & {\bar{\tilde z}}[1..N-1]^2 & {\bar{\tilde z}}[1..N-1]^N 
\end{array} \right)
\label{suNmat}
\eea 
Thus we immediately see that z[1], z[2],....,z[N-1] with (\ref{oncn}) 
describe SU(N) manifold. 
We will now exploit this simple characterisation 
of SU(N) manifold to construct coherent states belonging to 
all the irreducible representations of SU(N) group. For this 
purpose, we construct a new set of (N-1) parameters:
\bea 
z[1,1]^{i_{1}}~~ & \equiv & ~~z[1]^{i_{1}} \nonumber \\
z[1,2]^{i_{1}i_{2}} & \equiv & \sqrt{1 \over 2!} \sum_{p \in S_{2}} (-)^{|p|} 
z[1]^{i_{p_{1}}}z[2]^{i_{p_{2}}} \nonumber \\
z[1,3]^{i_{1}i_{2}i_{3}} &\equiv & \sqrt{1 \over 3!} \sum_{p \in S_{3}} (-)^{|p|} 
z[1]^{i_{p_{1}}}z[2]^{i_{p_{2}}} z[3]^{i_{p_{3}}} \nonumber \\
&...& \nonumber \\ 
z[1,N-1]^{i_{1}i_{2}..i_{N-1}} &\equiv & \sqrt{1 \over (N-1)!} 
\sum_{p \in S_{N-1}} (-)^{|p|} 
z[1]^{i_{p_{1}}}z[2]^{i_{p_{2}}}.. z[N-1]^{i_{p_{N-1}}} 
\label{np}
\eea 
\noindent We now define the SU(N) coherent states generating function as:
 \bea 
|z[1],z[2],..,z[N-1]>_{C_{1},C_{2}..C_{N-1}} \equiv
 exp\big(\sum_{\beta=1}^{N-1} z[1,\beta].a^{\dagger}[\beta] 
\big) |0>  
{\mid}_{{}^{a^{\dagger}[\alpha].a[\alpha] = C_{\alpha}}}
\nonumber \\
\equiv exp\Big(z[1]^{i^{1}_{1}}a^{\dagger}[1]_{i^{1}_{1}}  
+ z[1,2]^{i^{2}_{1}i^{2}_{2}} a^{\dagger}[2]_{i^{2}_{1}i^{2}_{2}}+...... 
\hspace{5cm} 
\nonumber \\
......+ z[1,N-1]^{i^{N-1}_{1}i^{N-1}_{2}...i^{N-1}_{N-1}} 
a^{\dagger}[N-1]_{i^{N-1}_{1}i^{N-1}_{2}..i^{N-1}_{N-1}}\Big) |0>  
{\mid}_{{}^{a^{\dagger}[\alpha].a[\alpha] = C_{\alpha}}}
\label{csn}
\eea 
The constraints on the right hand side of the generating function 
(\ref{csn}) ensure that we only generate the states which are 
eigenstates of all the (N-1) casimir operators and thus selecting
 all possible irreducible representations of SU(N), i.e, 
\bea 
{\cal{C}}[\alpha] |z[1],z[2],..,z[N-1]>_{C_{1},C_{2}..C_{N-1}} =
C[\alpha] |z[1],z[2],..,z[N-1]>_{C_{1},C_{2}..C_{N-1}},  
\label{eve} 
\eea  
for $\alpha = 1,2 ..(N-1)$. We will now show that the states thus obtained 
are the coherent states.  All states in (\ref{csn}) are defined continuously 
over the SU(N)  manifold parametrised by (z[1],z[2],...,z[N-1]). 
It is easy to check the SU(N) transformation properties of $z[1,\alpha],  
\alpha =1,2..(N-1)$. 
\bea 
z[1,\alpha]^{i_{1}..i_{\alpha}} & \rightarrow &  
z^{\prime}[1,\alpha]^{i_{1}..i_{\alpha}} = \Big(exp i\sum_{a=1}^{N^{2}-1} 
\theta^{a} 
\lambda^{a}[\alpha] \Big)^{i_{1}..i_{\alpha}}_{{\phantom{i_{1}..i_{\alpha}}}
j_{1}..j_{\alpha}} z[1,\alpha]^{j_{1}..j_{\alpha}} 
\label{transfn} 
\eea 
In (\ref{transfn}) $\theta^{a}$ are $(N^{2}-1)$ parameters describing a 
point on the SU(N) manifold.  The equation (\ref{transfn}) implies that 
all $z[\alpha]$ $(\alpha =1,2,..,N-1)$ transform as N-plets of SU(N), 
i.e,: 
\bea 
z[\alpha] \rightarrow z^{\prime}[\alpha] = \Big(exp i\sum_{a=1}^{N^{2}-1}
\theta^{a}
\lambda^{a}[1] \Big)^{i_{1}}_{
{\phantom{i_{1}}}
j_{1}} z[\alpha]^{j_{1}} 
\label{tr} 
\eea 
We see that the orthonormality conditions 
(\ref{onc}) remain invariant under the SU(N) 
transformations. The coherent states defined at a point (z[1],z[2]..
,z[1,N-1]) transform to the coherent states at $(z^{\prime}[1],z^{\prime}[2], 
.z^{\prime}[1,N-1])$ of the SU(N) manifold. 

\noindent Expanding the exponential we find: 
\bea 
|z[1],..,z[N-1]>_{C_{1},..C_{N-1}} = P \Big[\prod_{h=1}^{h_{max}}
\prod_{v=1}^{(C_{h} +C_{h+1}..+C_{N-1})} z^{i^{v}_{h}}[h]\Big] 
\Big[\prod_{v(1)=1}^{C_{N-1}} a^{\dagger}[N-1]_{i_{1}^{v(1)}i_{2}^{v(1)}....
i_{N-1}^{v(1)}}\Big] ~~~~~~\nonumber \\  
\Big[\prod_{v(2)=C_{N-1}+1}^{C_{N-1}+C_{N-2}} a^{\dagger}[N-2]_{i_{1}^{v(2)}i_{2}^{v(2)}....i_{N-2}^{v(2)}}\Big]
.....  \Big[\prod_{v(N-1)=C_{N-1}+..+C_{2}+1}^{C_{N-1}+...+C_{1}} 
a^{\dagger}[1]_{i_{1}^{v(N-1)}}\Big]|0> ~~~~~~~~~~~~
\label{csn21} 
\eea
In (\ref{csn21}), we have chosen those terms in the expansion which 
are eigenvectors of the Casimir operators ${\cal{C}}_{\beta}$ with 
eigenvalues $C_{\beta}$ and  $h_{max}$ is the maximum value\footnote{ 
In (\ref{csn21})  $h=1,2..h_{max}$ counts the horizontal rows and 
v is used to count the vertical columns of $\Sigma_{[C_{1},..,C_{N-1}]}$. 
Note that it is the maximum number of horizontal rows in the Young 
diagram.} 
of $\beta$ such that $C_{\beta} \neq 0$. $P = [\prod_{\alpha=1}^{N-1} 
P_{\alpha}]$ with $P_{\alpha} = (\alpha!)^{C_{\alpha} \over 2}$.  
We note the following important 
symmetry and anti-symmetry properties of (\ref{csn21}): 
\begin{enumerate} 
\item The (z[1],z[2],...z[N-1])  dependent part of (\ref{csn21}) has 
inbuilt invariance under interchange of any two indices along a row 
of the Young tableau $\Sigma_{[{C_{1},C_{2},....C_{N-1}}]}$, i.e, it 
is invariant under all the symmetry operations given in (\ref{symm}), 
\bea 
\Big(\prod_{h=1}^{h_{max}}\prod_{v=1}^{(C_{h} +C_{h+1}..
+C_{N-1})} z^{i^{v}_{h}}[h] \Big) {\cal{S_{\alpha}}} 
= \Big(\prod_{h=1}^{h_{max}}\prod_{v=1}^{(C_{h} +C_{h+1}..+C_{N-1})} z^{i^{v}_{h}}[h]\Big) 
\eea 
\item The harmonic oscillator dependent part of (\ref{csn21}) has 
inbuilt anti-symmetric properties along any column indices of 
the Young tableau $\Sigma_{[{C_{1},C_{2},....C_{N-1}}]}$. 
\end{enumerate} 
\noindent Therefore, we can write (\ref{csn21}) as: 
\bea
|z[1],..,z[N-1]>_{C_{1},..C_{N-1}}  =  \prod_{h=1}^{h_{max}}
\prod_{v=1}^{(C_{h} +C_{h+1}..+C_{N-1})} z^{i^{v}_{h}}[h]
~~~e_{\Sigma_{[C_{1},C_{2},...,C_{N-1}]}} \nonumber \\ 
 \prod_{v(1)=1}^{C_{N-1}} a^{\dagger}[N-1]_{i_{1}^{v(1)}i_{2}^{v(1)}....
i_{N-1}^{v(1)}}\prod_{v(2)=C_{N-1}+1}^{C_{N-1}+C_{N-2}} 
a^{\dagger}[N-2]_{i_{1}^{v(2)}i_{2}^{v(2)}....i_{N-2}^{v(2)}}
..\nonumber \\
...\prod_{v(N-1)=C_{N-1}+..C_{2}+1}^{C_{N-1}+...+C_{1}} a^{\dagger}
[1]_{i_{1}^{v(N-1)}}|0>
\label{csf11}
\eea 

\noindent In (\ref{csf11}), $e_{\Sigma_{[C_{1},C_{2},...,C_{N-1}]}}$ 
is defined by the equation (\ref{idemp}). Infact, these are exactly
basis vectors of the irreducible representation $\Sigma_{[C_{1},C_{2}.
.......,C_{N-1}]}$. We denote them by: 
\bea 
\Vert (i_{1}^{1} i_{1}^{2} .......i_{1}^{C_{1}+..
+C_{N-1}}); (i_{2}^{1} i_{2}^{2} .......i_{2}^{C_{2}+..+
C_{N-1}}); .......;(i_{N-1}^{1} i_{N-1}^{2} .......
i_{N-1}^{C_{2}+..+C_{N-1}})  \rangle_{\Sigma_{[C_{1},C_{2},...,C_{N-1}]}}  
 \nonumber \\
\equiv e_{\Sigma_{[C_{1},C_{2},...,C_{N-1}]}} 
\prod_{v(1)=1}^{C_{N-1}} a^{\dagger}[N-1]_{i_{1}^{v(1)}i_{2}^{v(1)}....
i_{N-1}^{v(1)}}
\prod_{v(2)=C_{N-1}+1}^{C_{N-1}+C_{N-2}}
a^{\dagger}[N-2]_{i_{1}^{v(2)}i_{2}^{v(2)}....i_{N-2}^{v(2)}} \nonumber \\
.....................  \prod_{v(N-1)=C_{N-1}+..C_{2}+1}^{C_{N-1}+...
+C_{1}} a^{\dagger}[1]_{i_{1}^{v(N-1)}}|0> {\hspace{4cm}}
\label{irrepnn} 
\eea

\noindent Thus the states belonging to $\Sigma_{[C_{1},..,C_{N-1}]}$ 
in (\ref{csn}) irreducible representations are given by: 
\bea 
\label{uud} 
|z[1],z[2],..,z[N-1]>_{C_{1},C_{2}..C_{N-1}} = 
\Big(\prod_{h=1}^{h_{max}}\prod_{v=1}^{(C_{h} +
C_{h+1}..+C_{N-1})} z^{i^{v}_{h}}[h]\Big)   {\hspace{3cm}} \\
\Vert (i_{1}^{1} i_{1}^{2} .......i_{1}^{C_{1}+..
+C_{N-1}}); (i_{2}^{1} i_{2}^{2} .......i_{2}^{C_{2}+..+
C_{N-1}}); .......;(i_{N-1}^{1} i_{N-1}^{2} .......
i_{N-1}^{C_{2}+..+C_{N-1}}) 
\rangle_{\Sigma_{[C_{1},C_{2},...,C_{N-1}]}}   \nonumber 
\eea
\noindent We now check the resolution of identity. This is similar 
to the SU(3) case in the previous sections.  The SU(N) Haar measure 
is: 
\bea
\int d\mu (z) \equiv  \Big(\int\prod_{\alpha=1}^{N-1}dz[\alpha]\Big)
\Big(\prod_{\alpha,\beta}\delta(z[\alpha].z^{*}[\beta]
-\delta_{\alpha,\beta})\Big)
\eea
\noindent We construct an operator ${\cal{O}}_{[N]}$:
\bea
{\cal{O}}_{N} \equiv \Big(\int d\mu(z)
|z[1],z[2],..,z[N-1]>_{C_{1},C_{2}..C_{N-1}}{~~}_{C_{1},C_{2}..C_{N-1}}
<z[1],z[2],..,z[N-1]|
\label{opn}
\eea
\noindent Under SU(N) transformations (\ref{transfn}), ${\cal{O}}_{[3]}$
remains invariant. Therefore,
\bea
[Q^{a},{\cal{O}}_{[N]}] = 0, ~~~~ \forall a =1,2,....,N^{2}-1
\eea
\noindent The Schur's Lemma implies:
\bea
{\cal{O}}_{[N]} = K I_{[C_{1},C_{2},...,C_{N-1}]}
\label{idn}
\eea
\noindent In (\ref{idn}), K is a constant and
$I_{[C_{1},..,C_{N-1}]}$ is the identity operator in
the $\Sigma_{[C_{1},..,C_{N-1}]}$  irreducible
representation subspace. Therefore, the states in 
(\ref{uud}) are indeed the coherent states.  
We have thus constructed all 
the SU(N) coherent states  belonging to  
different irreducible representations of SU(N).

\section{Summary and Discussion} 

We have generalized Schwinger representation of SU(2) algebra in terms 
of harmonic oscillotrs to the group SU(N). We have exploited this 
construction to construct SU(N) coherent states and characterized SU(N) 
manifold in terms of complex vectors. In this sense our SU(N) coherent states 
definition is analogous to that of Heisenberg-Weyl coherent states. 
This method is quite general and can be generalized to other Lie groups 
and  their manifolds.  
We feel our approach is more useful for practical calculations 
compared to the standard group theoretical approach and can be 
applied to various problems. 
In condensed matter physics, coherent states for the
Lie group $SU(2)$ have been very useful for studying Heisenberg spin systems
using the path integral formalism \cite{arovas,manous,fradkin,sachdev}.
These studies have been generalized to systems with $SU(N)$ symmetry, although
such studies have usually been restricted to the completely symmetric
representations \cite{manous,sun}.  Therefore, our formulation can be used 
to write down the field theory for the 
$SU(N)$ Heisenberg model and study its spectrum and topological aspects  
as in the $SU(2)$ case \cite{haldane}. Using the techniques discussed 
in the paper, one can also construct SU(N) coherent states with fixed 
values of Cartan diagonal generators. The special cases of SU(2) 
coherent states with fixed charge and SU(3) nonlinear coherent states 
with fixed charge and hypercharge were constructed in \cite{bdr} and 
\cite{fan} respectively.  

\begin{flushleft} 
{\bf Acknowledgments} 
\end{flushleft} 

\noindent We would like to thank Prof. Binayak Dutta Roy for useful 
discussions. 
One of the authors (M.M) would also like to thank Prof. Diptiman Sen 
for his involvement and discussions at the initial stages of this work. 
\vskip .8 true cm

\end{document}